\documentstyle[12pt]{article}
\input epsf
\begin{document}
\title{ 
\hfill hep-ph/0007334\\ \vskip .5truecm
{\large {\bf Phenomenology of Pseudo Dirac Neutrinos}}} 
\vskip 2.0truecm
\author{ Anjan S. Joshipura$^{1,2}$ and Saurabh D. Rindani$^{1,3}$\\
{\ns\it $^1$Theoretical Physics Group, Physical Research Laboratory,}\\
{\ns\it Navarangpura, Ahmedabad, 380 009, India.}\\[.1cm]
{\ns \it $^2$Theory Division, CERN CH-1211, Geneva 23,
Switzerland}\\[.1cm]
{\ns \it$^3$Theory Group, DESY, 22603, Hamburg, Germany}}
\date{}
\def\ap#1#2#3{           {\it Ann. Phys. (NY) }{\bf #1} (19#2) #3}
\def\arnps#1#2#3{        {\it Ann. Rev. Nucl. Part. Sci. }{\bf #1} (19#2) #3}
\def\cnpp#1#2#3{        {\it Comm. Nucl. Part. Phys. }{\bf #1} (19#2) #3}
\def\apj#1#2#3{          {\it Astrophys. J. }{\bf #1} (19#2) #3}
\def\asr#1#2#3{          {\it Astrophys. Space Rev. }{\bf #1} (19#2) #3}
\def\ass#1#2#3{          {\it Astrophys. Space Sci. }{\bf #1} (19#2) #3}

\def\apjl#1#2#3{         {\it Astrophys. J. Lett. }{\bf #1} (19#2) #3}
\def\ass#1#2#3{          {\it Astrophys. Space Sci. }{\bf #1} (19#2) #3}
\def\jel#1#2#3{         {\it Journal Europhys. Lett. }{\bf #1} (19#2) #3}

\def\ib#1#2#3{           {\it ibid. }{\bf #1} (19#2) #3}
\def\nat#1#2#3{          {\it Nature }{\bf #1} (19#2) #3}
\def\nps#1#2#3{          {\it Nucl. Phys. B (Proc. Suppl.) }
                         {\bf #1} (19#2) #3} 
\def\np#1#2#3{           {\it Nucl. Phys. }{\bf #1} (19#2) #3}
\def\pl#1#2#3{           {\it Phys. Lett. }{\bf #1} (19#2) #3}
\def\pr#1#2#3{           {\it Phys. Rev. }{\bf #1} (19#2) #3}
\def\prep#1#2#3{         {\it Phys. Rep. }{\bf #1} (19#2) #3}
\def\prl#1#2#3{          {\it Phys. Rev. Lett. }{\bf #1} (19#2) #3}
\def\pw#1#2#3{          {\it Particle World }{\bf #1} (19#2) #3}
\def\ptp#1#2#3{          {\it Prog. Theor. Phys. }{\bf #1} (19#2) #3}
\def\jppnp#1#2#3{         {\it J. Prog. Part. Nucl. Phys. }{\bf #1} (19#2) #3}

\def\rpp#1#2#3{         {\it Rep. on Prog. in Phys. }{\bf #1} (19#2) #3}
\def\ptps#1#2#3{         {\it Prog. Theor. Phys. Suppl. }{\bf #1} (19#2) #3}
\def\rmp#1#2#3{          {\it Rev. Mod. Phys. }{\bf #1} (19#2) #3}
\def\zp#1#2#3{           {\it Zeit. fur Physik }{\bf #1} (19#2) #3}
\def\fp#1#2#3{           {\it Fortschr. Phys. }{\bf #1} (19#2) #3}
\def\Zp#1#2#3{           {\it Z. Physik }{\bf #1} (19#2) #3}
\def\Sci#1#2#3{          {\it Science }{\bf #1} (19#2) #3}
\def\n.c.#1#2#3{         {\it Nuovo Cim. }{\bf #1} (19#2) #3}
\def\r.n.c.#1#2#3{       {\it Riv. del Nuovo Cim. }{\bf #1} (19#2) #3}
\def\sjnp#1#2#3{         {\it Sov. J. Nucl. Phys. }{\bf #1} (19#2) #3}
\def\yf#1#2#3{           {\it Yad. Fiz. }{\bf #1} (19#2) #3}
\def\zetf#1#2#3{         {\it Z. Eksp. Teor. Fiz. }{\bf #1} (19#2) #3}
\def\zetfpr#1#2#3{       {\it Z. Eksp. Teor. Fiz. Pisma. Red. }{\bf #1} (19#2) #3}
\def\jetp#1#2#3{         {\it JETP }{\bf #1} (19#2) #3}
\def\mpl#1#2#3{          {\it Mod. Phys. Lett. }{\bf #1} (19#2) #3}
\def\ufn#1#2#3{          {\it Usp. Fiz. Naut. }{\bf #1} (19#2) #3}
\def\sp#1#2#3{           {\it Sov. Phys.-Usp.}{\bf #1} (19#2) #3}
\def\ppnp#1#2#3{           {\it Prog. Part. Nucl. Phys. }{\bf #1} (19#2) #3}
\def\cnpp#1#2#3{           {\it Comm. Nucl. Part. Phys. }{\bf #1} (19#2) #3}
\def\ijmp#1#2#3{           {\it Int. J. Mod. Phys. }{\bf #1} (19#2) #3}
\def\ic#1#2#3{           {\it Investigaci\'on y Ciencia }{\bf #1} (19#2) #3}
\def\tp{these proceedings}
\def\pc{private communication}
\def\ip{in preparation}
\relax

\newcommand{\GeV}{\,{\rm GeV}}
\newcommand{\MeV}{\,{\rm MeV}}
\newcommand{\keV}{\,{\rm keV}}
\newcommand{\eV}{\,{\rm eV}}
\newcommand{\Tr}{{\rm Tr}\!}
\renewcommand{\arraystretch}{1.2}
\newcommand{\beq}{\begin{equation}}
\newcommand{\eeq}{\end{equation}}
\newcommand{\beqa}{\begin{eqnarray}}
\newcommand{\eeqa}{\end{eqnarray}}
\newcommand{\ba}{\begin{array}}
\newcommand{\ea}{\end{array}}
\newcommand{\bmat}{\left(\ba}
\newcommand{\emat}{\ea\right)}
\newcommand{\refs}[1]{(\ref{#1})}
\newcommand{\ler}{\stackrel{\scriptstyle <}{\scriptstyle\sim}}
\newcommand{\ger}{\stackrel{\scriptstyle >}{\scriptstyle\sim}}
\newcommand{\lag}{\langle}
\newcommand{\rag}{\rangle}
\newcommand{\ns}{\normalsize}
\newcommand{\cm}{{\cal M}}
\newcommand{\gr}{m_{3/2}}
\newcommand{\p}{\partial}

\def\rp{ $R_P$} 
\def\321{$SU(3)\times SU(2)\times U(1)$}
\def\tl{{\tilde{l}}}
\def\tL{{\tilde{L}}}
\def\bd{{\overline{d}}}
\def\tL{{\tilde{L}}}
\def\a{\alpha}
\def\b{\beta}
\def\g{\gamma}
\def\c{\chi}
\def\d{\delta}
\def\D{\Delta}
\def\db{{\overline{\delta}}}
\def\Db{{\overline{\Delta}}}
\def\e{\epsilon}
\def\l{\lambda}
\def\n{\nu}
\def\m{\mu}
\def\nt{{\nu_{\tau}}}
\def\p{\phi}
\def\P{\Phi}
\def\x{\xi}
\def\r{\rho}
\def\s{\sigma}
\def\t{\tau}
\def\th{\theta}
\def\ne{\nu_e}
\def\nm{\nu_{\mu}}
\def\rp{$R_P$}
\def\mp{$M_P$}
\def\mgut{M_{GUT}}
\def\emt{$L_e-L_\m-L_{\tau}$ }     
\renewcommand{\Huge}{\Large}
\renewcommand{\LARGE}{\Large}
\renewcommand{\Large}{\large}
\maketitle
\vskip 2.0truecm
\begin{abstract}
We formulate general conditions  on $3\times 3$ neutrino mass matrices  
under which
a degenerate pair of neutrinos at a high scale would split at low scale 
by radiative corrections involving only the standard model fields.
This generalizes the original observations of Wolfenstein on pseudo Dirac 
neutrinos
to three generations. A specific model involving partially broken discrete
 symmetry
and solving  the solar and
atmospheric 
anomalies is proposed. The symmetry pattern of the model naturally
generates two large angles one of which can account for the large angle
MSW solution to the solar neutrino problem. \end{abstract}
\section{Introduction}
Dirac neutrinos are associated with an unbroken U(1) symmetry acting on
leptons. A small breaking of this symmetry splits a Dirac neutrino into
a pair of majorana neutrinos with  
(mass)$^2$ difference much smaller than the square of the original mass.
Such a pair can simultaneously describe small splitting and large mass
and is of phenomenological importance in solving solar and/or atmospheric
 neutrino
anomalies \cite{smir}.

Zeroth order approximation to a pseudo Dirac neutrino is provided by the 
following texture in case of two generations say, $e$ and $\m$
\beq
\label{ps1}
\left( \ba{cc} 0&m\\m&0\\\ea \right) \eeq 
This displays an unbroken $L_e-L_\m$ symmetry and its breaking ( by introducing
small non-zero diagonal elements) leads to a pseudo Dirac neutrino. This
breaking can be 
explicitly introduced by allowing for additional fields like Higgs triplet or
right handed neutrinos \cite{pdt} or it can be introduced
radiatively by breaking the symmetry $L_e-L_\m$ in the charged lepton sector
\cite{rad1,rad2}.

A slightly non-trivial example of the pseudo Dirac neutrino is provided by 
a mass matrix discussed by Wolfenstein \cite{wolf}
\beq
\label{ps2}
\left( \ba{cc} a&b\\b&-a\\ \ea \right) \eeq

Both the textures in eqs.(1) and (2) lead to a pair of neutrinos with equal and 
opposite eigenvalues. However they differ conceptually and phenomenologically
from each other \footnote{ We are assuming here that these textures are
defined in a basis with a diagonal charged lepton mass matrix.}. 
Unlike in eq.(1),
the matrix in eq.(2) cannot be invariant under a $U(1)$ symmetry 
corresponding to 
{\it any} combination of 
lepton numbers. This has the consequence that the charged current interactions
defined in the mass basis of the degenerate pairs violate 
lepton number \cite{wolf}
and the 
pseudo Dirac pair gets split automatically by radiative
corrections \cite{petcov}. 
Thus the theory described by eq. (2)
intrinsically defines a pseudo Dirac neutrino while one needs to invoke 
additional 
fields in order to break the $L_e-L_\m$ symmetry in case of eq.(1). At the
phenomenological level, the mixing implied by (1) is maximal 
while it is arbitrary in case (2). 
The most phenomenological discussions related to pseudo Dirac neutrinos
in the literature \cite{pdt} are in the context of  texture in (1). We wish 
to discuss
here instead several interesting aspects related to the Wolfenstein
 texture, eq.(2) and its generalization to three families.

The radiative splitting of neutrinos in case (2) is most simply demonstrated
using the relevant renormalization group (RG) equations. Assume that the
neutrino mass matrix in (\ref{ps2}) is specified at some high scale $M_X$ 
and the effective 
theory below this scale is the standard model (SM) or the minimal 
supersymmetric standard model (MSSM). The neutrino mass matrix at a 
low scale $\m\sim M_Z$ is then given by \cite{el}
\beq
\label{rg}
M_\n(\m)\approx P M_\n(M_X) P\;, \eeq
where in the three-generation case $P=Diag.(1+\d_e,1+\d_\m,1+\d_\t)$ and  
\beq
\label{deltas}
\d_\a\equiv - {m_\a^2\over v^2\cos^2\b(4 \pi)^2}\ln{M_X\over \mu}
\eeq
in case of the minimal supersymmetric standard model. $m_\a,v\sim 174
\GeV$ here refer to the charged lepton masses and the weak scale
respectively.
As before, $M_\n$ is specified in the physical basis of the charged 
leptons. The texture (2) at high scale gets transformed to the texture
\beq
\label{ps2m}
\left( \ba{cc} a(1+2\d_e)&b(1+\d_e+\d_\m)\\b(1+\d_e+\d_\m)&-a (1+2 \d_\m) \\ 
\ea \right)+O(\d^2) \eeq
at the low scale. This describes a split pair of neutrinos. In contrast,  
eq.(1) leads to a degenerate pair even after RG evolution as in eq.(\ref{rg})
is taken into account. 

The modified mixing angle and the mass splitting implied by eq.(\ref{ps2m}) 
are given by
\beqa 
\label{split2}
\tan 2\theta(\m)&\approx&\tan 2\theta(M_X) (1 +O(\d^2))\;, \nonumber \\
\Delta &\approx& 4 m_0^2 \cos 2 \theta \d_\m \; .\eeqa
with $m_0=\sqrt{a^2+b^2}$ and $\tan 2\theta={b\over a}$.
As follows from eq.(\ref{deltas}), typical strength of radiative corrections
is $\d_\m\sim 10^{-7}$. From phenomenological point of view, the required
 value of
$\Delta$ can be as small
as $10^{-11}\eV^2$. If one starts with a mass matrix having such splitting
at $M_X$ then there is a possibility that the radiative corrections may lead to
much larger splitting than this. Likewise, maximal mixing angle at $M_X$
can also get 
destabilized \cite{el}. In the present case, splitting is zero at $M_X$ and 
it is only induced by
radiative corrections. As follows from eq.(\ref{split2}), the magnitude of this
splitting can be
in the range $10^{-10}-10^{-11} \eV^2$ for $m_0$ near the atmospheric
neutrino scale. Moreover, 
the mixing angle at $M_X$ is arbitrary and 
receives corrections only at O($\d^2$) in this case. Thus, the texture in (2)
 is stable against radiative corrections unlike some of the textures 
discussed in \cite{el}. These properties  make the Wolfenstein texture in
(\ref{ps2})
also phenomenologically realistic. Theoretically, this 
texture is not a fine tuned possibility but can arise from imposition of 
the following discrete symmetry
on neutrino mass matrix:  

\beq
\label{sym1}
L_e\rightarrow i L_\m\; ; \;\;\;\;\;\;\; L_\m\rightarrow -iL_e \eeq
This symmetry is broken by hierarchical charged lepton masses which lead to 
radiative splitting between the degenerate pair.
 
The $2\times 2$ texture of eq.(\ref{ps2}) is successful but not complete 
from the point of view of simultaneous solution to the solar and atmospheric
neutrino anomalies. This would require going beyond two generations. 
The purpose of this note is to generalize the above considerations to
the realistic case of three generations and identify phenomenologically viable 
models/textures leading to pseudo Dirac neutrinos. We first write down the 
general
conditions on an arbitrary
$3\times 3$ neutrino mass matrix under which it leads to a pair of
degenerate neutrinos.
Such matrices can be classified in two categories, those in which degeneracy is
preserved by radiative corrections involving standard model fields 
and those in which the theory describes a 
pseudo Dirac state. We formulate general criteria to distinguish between these
two categories 
and show that they  can be identified by looking at
the structure of the leptonic mixing matrices implied at the
{\it tree level}. Then we discuss an example which satisfies 
phenomenological requirements to obtain a solution to solar and
atmospheric neutrino anomalies. Starting with a generalization of the
original Wolfenstein mass matrix,  splitting of neutrino states needed for
solar neutrino anomaly arises in this example through radiative
corrections. The model can lead to either vacuum solution with bi-maximal
mixing or MSW solution corresponding to large mixing angle.
\section{Pseudo Dirac neutrinos: General analysis}
Let us consider a CP conserving theory specified by a general $3\times 3$
real symmetric mass matrix $M_\n$ for the neutrinos:
\beq
\label{mass1}
-{\cal L}_m={1\over 2} \overline{({\nu'}_{\a L})^c} (M_\n)_{\a\b}\nu'_{\b L}
+H. c. \; . \eeq
The $M_\n$ would contain two equal and opposite eigenvalues if it
satisfies the following condition:
\beq
\label{condtree}
tr(M_\n)\sum_{i}\Delta_i=det M_\n\; , \eeq
where $\Delta_i$ represents the determinant of the $2\times 2$ block of $M_\n$
obtained by blocking $i^{th}$ (i=1,2,3) row and column.
Only $M_\n$ satisfying condition (\ref{condtree}) would lead to a 
Dirac or pseudo Dirac neutrinos.
Define a $U^\n$  which diagonalizes such $M_\n$:
\beq
\label{mnud}
U^\n\; M_\n U^{\n T}=Diag. (m, -m, m')\; .
\eeq
The physical mass basis for neutrinos is defined by $\nu_{L}=U^{\n}\nu'_L$.
The neutrino masses can be written in terms of Majorana spinors 
\beq 
\nu_{1,3}=\nu_{1,3 L}+(\nu_{1,3 L})^c\;\;\;\;\; \nu_2=\nu_{2 L}-(\nu_{2 L})^c 
\; .\eeq
Explicitly,
\beq
\label{mass2}
-{\cal L}_m={1\over 2} \left[m (\bar{\nu}_{1}\nu_{1}+\bar{\nu}_{2}\nu_{2})
+m' \bar{\nu}_{3}\nu_{3}\right]\; .\eeq 
The degeneracy of two of the Majorana states allows us to define
\beq 
\psi={1\over \sqrt{2}}(\nu_{1}+\nu_2) \eeq
and rewrite eq.(\ref{mass2}) as
\beq
-{\cal L}_m= m \bar{\psi}\psi+{1\over 2}m' \bar{\nu_3}\nu_3 \; .\eeq
Note that the $\psi$ is a four component Dirac field since 
$\psi^c \not = \psi$.
As a result, the system specified  by eq.(\ref{condtree}) corresponds 
to a Dirac and a majorana neutrino.

The charged current interactions can be written as follows in the 
leptonic mass basis: 
\beqa
\label{ch1}
-{\cal L}_W&=&{ g\over \sqrt{2}}\bar{e'}_{\a L}\g_\m\nu'_{\a L}W^\m+ H.
c.\;,
\nonumber \\
&=&{ g\over \sqrt{2}}\bar{e}_{\a L}\g_\m\left( {1\over
\sqrt{2}}(K_{\a1}+K_{\a 2})
\psi_L+
{1\over \sqrt{2}}(K_{\a1}-K_{\a 2})(\psi^c)_L+ K_{\a 3}\nu_{3 L}\right)W^\m
+H.c. \; .\nonumber \\
\eeqa
Here $e_{a L}$ represents the physical mass basis for the charged leptons and
the $K$ represents the leptonic Kobayashi Maskawa (KM) matrix.

Eq.(\ref{ch1}) is a straightforward generalization of the $2\times 2$ case 
considered in 
\cite{wolf}. It shows that although mass term for $\psi$ is invariant under
 a U(1) symmetry the charged current violates it and the Dirac state will
split 
by radiative corrections \cite{petcov}. Thus any $3\times 3 $ matrix
satisfying condition 
(\ref{condtree})
at tree level would  generically lead to a pseudo Dirac state.

While the lepton number violation is generically present, it is easy to identify
all the $3\times 3$ structures for $M_\n$ which admit  an unbroken $U(1)$
symmetry 
corresponding  to a truly Dirac 
neutrino. Necessary condition for this to happen is easy to write down 
using eq.(\ref{ch1}):
\beq
\label{cond1}
K_{\a 1}=\eta_\a K_{\a 2}  \eeq
Here $\eta_\a=\pm 1$. The above equation ensures that either only $\psi$ or 
$\psi^c$ couples to a given charged lepton $e_\a$. As a result, phase rotation
of $\psi$ can be symmetry of eq.(\ref{ch1}). 

The above condition is also sufficient to ensure truly Dirac state in case 
of two generations. But the couplings of the third generation to the first
two requires additional constraint to obtain a $U(1)$ symmetry.
To see this, note that 
orthogonality of $K$ does not allow all the $\eta_a$ in eq.(\ref{cond1})
to have the same sign. Without loss of generality we can choose 
$\eta_1=-\eta_2=-\eta_3=1$. Other solutions of eq.(\ref{cond1}) are obtained by
interchange $1\leftrightarrow 2,1\leftrightarrow 3$ or an overall
multiplication of  $\eta_a$ by -1 in all these cases. 
It is easy to write a parameterization of  $K$ with this 
choice of $\eta$ using orthogonality.
\beq
\label{genk}
K=
\left(
\ba{ccc}
{1\over \sqrt{2}}&{1\over \sqrt{2}}&0\\
{-c\over \sqrt{2}}&{c\over \sqrt{2}}&s\\
{s\over \sqrt{2}}&{-s\over \sqrt{2}}&c\\ \ea \right) \eeq

Apart from interchange of rows and overall multiplications of any column by -1, 
this is the most general form for any real $K$ satisfying eq.(\ref{cond1}). 
Using above eq. (\ref{genk})
in (\ref{ch1}) we immediately see that if
$s=0$, the third neutrino does not mix with the first two and eq.(\ref{ch1}) is 
invariant under the symmetry,
$$ \psi_L\rightarrow e^{i\a} \psi_L ;\;\;\;\;\; e_L\rightarrow e^{i\a} e_L\;;\;\;\;\;
\m_L\rightarrow e^{-i\a}\mu_L $$
In this case, the third neutrino is decoupled and we get an unbroken $L_e-L_\m$
symmetry and a truly Dirac neutrino.

Even when $s\not =0$, $W$ interactions can still be made invariant under the 
following symmetry
$$
\psi_L\rightarrow e^{i\a} \psi_L ;\;\;\;\;\; e_L\rightarrow e^{i\a} e_L\;;\;\;\;\;
(\m_L,\tau_L,\nu_{3L})\rightarrow e^{-i\a}(\mu_L,\tau_L,\nu_{3L}) $$
But this symmetry would be violated by the mass of the third neutrino
and hence we would expect splitting of the neutrino pairs radiatively in 
this case even though the 
charged current interactions possess an unbroken symmetry.  

It follows from the above discussion that there are two ways in which a 
degenerate pair at high scale would be split by the radiative corrections
involving 
$W$: (i) When mixing matrix does not satisfy the necessary condition in
(\ref{cond1}) and (ii) when mixing matrix satisfies this condition but the 
third
neutrino has a mass and is not decoupled from the first two. The latter
situation is more interesting since in this
 case the implied form of $K$, eq.(\ref{genk}) corresponds to bi-maximal
 mixing which is argued \cite{bim,bim1} to be useful in simultaneous solution
 of the solar and atmospheric
neutrino anomalies. The mass of the third generation plays a non-trivial role
in splitting the degenerate pair in this case. This was found to be true in 
specific example
considered in \cite{brs}. The present discussion highlights the importance of
a non-zero third generation mass from more general considerations.

All the above conclusions were based on the $U(1)$ invariance of 
the $W$ interactions displayed in eq. (\ref{ch1}).
The same conclusions can be drawn
from the study of the RG evolution of the texture as in eq.(\ref{rg}).
The matrix $K$ in eq.(\ref{genk}) implies the following texture for 
the neutrino mass matrix in the charged lepton mass basis:
\beq
\label{gent}
\left( \ba{ccc}
0&-mc&ms\\
-mc&m's^2&m'cs\\
ms&m'cs&m'c^2 \\
\ea \right) \eeq
Assuming this texture to be true at $M_X$, texture at a lower scale
can be worked out using eq.(\ref{rg})
\beq
\label{gent2}
\left( \ba{ccc}
0&-mc(1+\d_e+\d_\m)&ms(1+\d_e+\d_\t)\\
-mc(1+\d_e+\d_\m)&m's^2(1+2\d_\m)&-m'cs(1+\d_\t+\d_\m)\\
ms (1+\d_e+\d_\t)&-m'cs(1+\d_\t+\d_\m)&m'c^2 (1+2 \d_\t)\\
\ea \right) \eeq
This texture is seen to lead to a degenerate pair at low scale 
if $s=0$ or $m'=0$ in accordance with the conditions discussed above.
Otherwise, it describes a pseudo Dirac state. In the former case, the 
neutrino mass
matrix in the charged lepton mass basis respects either $L_e-L_\m$ ($s=0$) 
or $L_e-L_\m-L_\t$ ($m'=0$) symmetry. In the latter case, splitting occurs but
only at O($\d^2$) as can be seen using eq.(\ref{gent2}). 
This makes the case (ii)
discussed above also more natural from the point of view of stability as
argued in \cite{brs}. 

\section{A Specific Model}

A very economical scheme for understanding the solar and atmospheric
neutrino anomaly is provided by the following scenario \cite{rad1,bim1}.
The
neutrino 
spectrum at tree level consists of  one massless neutrino and two 
degenerate neutrinos with mass in the atmospheric neutrino range.
The radiative corrections split this pair and provides the scale
needed to understand the solar neutrino anomaly. Different possibilities 
realizing this scenario have been proposed \cite{rad1,bim1}. We give here
a
specific
and very economical example in the context of pseudo Dirac texture
where one does not need to invoke any new physics and the $W$ interactions
provide a source for the solar scale.

As we indicated in introduction, the Wolfenstein type structure as
in eq.(\ref{ps2})  can
result from a  discrete symmetry. A similar symmetry can be used to
obtain a $3\times 3$ generalization of eq. (\ref{ps2}). Consider,
\beq
\label{sym2}
(L_e',L_\m')\rightarrow i (L_\m',L_e')\; ;\;\;\;\;\;\;\;\;
(L_\tau',e_R',\m_R',\t_R')\rightarrow -i (L_\tau',\m_R',e_R',\t_R')
\eeq
Here, $L'_\a$ and $(e'_R,\m'_R,\tau'_R)$ denote the leptonic doublets and
singlets respectively.
All other fields are assumed neutral with respect to the above symmetry.
The neutrino mass matrix invariant under this symmetry has the following form:
\beq
\label{mnu}
M_\n(M_X)=\left( \ba{ccc} 
a&0&b\\
0&-a&b\\
b&b&0\\
\ea \right)\eeq
This mass matrix satisfies condition, eq.(\ref{condtree}) and hence leads to 
a degenerate pair 
of neutrinos with mass $m=\sqrt{a^2+2 b^2}$ at the tree level. It is not in
 the basis
with diagonal charged leptons. The symmetry in eq.(\ref{sym2}) allows the 
following 
general form for the charged lepton masses:
\beq
\label{mcl}
M_l=\left( \ba{ccc} 
m_1&m_2&m_3\\
-m_2&-m_1&-m_3\\
m_4&m_4&m_5\\
\ea \right)\eeq
Here $m_i$ are parameters arising in the standard way through Yukawa
 couplings of
leptons to the Higgs field. The $M_l$ can be diagonalized in the standard 
manner.
Eq.(\ref{mcl}) leads to
\beq
\label{mlmld}
M_lM_l^{\dagger}=\left( \ba{ccc}
A&B&C\\
B&A&-C\\
C&-C&D\\ \ea \right) \eeq
The parameters in the above matrix can be read off from eq.(\ref{mcl}).
One finds,
\beq
U^l M_lM_l^{\dagger}U^{l\dagger}= Diag. (m_e^2,m_\m^2,m_\tau^2) \eeq
With 
\beq
\label{ul}
U^l=R_{23}(\phi) R_{12}(\pi/4)=
\left(
\ba{ccc}
{1\over \sqrt{2}}&{1\over \sqrt{2}}&0\\
{-c_\phi\over \sqrt{2}}&{c_\phi\over \sqrt{2}}&s_\phi\\
{s_\phi\over \sqrt{2}}&{-s_\phi\over \sqrt{2}}&c_\phi\\ \ea \right) \eeq
where $c_\phi=\cos\phi;\; s_\phi=\sin\phi$ and 
$$ \tan 2 \phi= {2\sqrt{2}C\over D-A+B}$$
While fairly large ranges can be allowed for 
the parameters $m_i$, we will make the following choice for
illustrative purpose:
\beq
\label{illu}
m_1\sim m_2\sim {1\over 2} m_e\; ;\;\;\;\;\ m_3\sim O(m_\m)\; ;\;\;\;\;\;
m_5\sim O(m_\t) \eeq
It can be seen that this choice reproduces the charged lepton masses correctly.
Moreover, with this choice,
\beq
\label{value}
\tan{2 \phi}\sim O({m_\m\over m_\tau}) \eeq

The neutrino mass matrix in eq.(\ref{mnu}) is diagonalized by
\beq
\label{unu}
U_\n=\left( \ba{ccc} 
\cos^2{\theta\over 2}&\sin^2{\theta\over 2}&\frac{1}{\sqrt{2}}\sin\theta\\
\sin^2{\theta\over 2}&\cos^2{\theta\over 2}&-\frac{1}{\sqrt{2}}\sin\theta\\
-\frac{1}{\sqrt{2}}\sin\theta&{1\over \sqrt{2}}\sin\theta&\cos\theta\\
\ea \right)\eeq
with $\tan\theta\equiv{\sqrt{2} b\over a}$.

In spite of the complicated form for 
$U_\n$, the KM matrix $K\equiv U^lU^{\nu T}$ has a bi-maximal form
given in eq.(\ref{genk}). This form coupled with the masslessness of the third
generation ensures that the degenerate pair does not split radiatively.
This is not
surprising since the charged lepton mass matrix, eq.(\ref{mcl}) is also
invariant under
the symmetry of eq.({\ref{sym2}) which was responsible for the
degeneracy of neutrinos.
We need to break this symmetry in order to obtain splitting of the degenerate 
pairs. One possible breaking is as follows:
\beq
\label{m'}
\tilde{m} (\bar{e'}_Le_R'+\bar{\m'}_L\m_R') \eeq
Such a term can arise from Yukawa couplings with the standard Higgs
in which case it breaks the symmetry explicitly. This type of hard breaking
advocated in several papers \cite{rad2}  can make the theory non-renormalizable. This is
avoided by Yukawa couplings of an additional
Higgs field which is odd under the discrete symmetry. The discreet symmetry
 would 
then be spontaneously broken. Alternatively, we can work with the hard breaking 
in (\ref{m'}) but
assume the model to be embedded in the supersymmetric theory which would not
make the model non-renormalizable. We shall choose the latter alternative
in what follows.

The inclusion of the term in eq.(\ref{m'}), would be expected to
split the degeneracy
between the neutrinos. This can be seen after some algebra. Eq.(\ref{mlmld}) 
is now replaced
by 
\beq
M_lM_l^{\dagger}=\left( \ba{ccc}
A'+\e_1&B&C+\e_2\\
B&A'-\e_1&-C+\e_2\\
C+\e_2&-C+\e_2&D\\ \ea \right) \eeq
with $A'=A+\tilde{m}^2$  , $\e_1=2
m_1 \tilde{m}$ and $\e_2=\tilde{m}m_4$. 
The above matrix can be diagonalized by the following $U^l$ if one assumes
$$ \tan\theta_{13}\approx {\e_1\sin\phi'+\sqrt{2}\e_2\cos\phi'\over
m_\tau^2}\ll 1$$

\beq
\label{ulm}
U^l=R_{12}(\theta_{12})R_{23}(\phi') R_{12}(\pi/4) \eeq
where 
\beqa
\tan 2\phi'&\approx& {2\sqrt{2}C'\over D-A'+B'};\nonumber \\
\tan 2\theta_{12}&\approx&{2 (\e_1\cos\phi'-\sqrt{2}\e_2\sin\phi')\over 
B(1+\cos^2\phi ')+ A'\sin^2\phi ' + \sqrt{2}C'\sin 2\phi '
-D\sin^2\phi '}
\eeqa

The KM matrix following from eqs.(\ref{unu},\ref{ulm}) is given by
\beq
\label{kmx}
K(M_X)=R_{12}(\theta_{12})R_{23}(\theta_{23}) R_{12}(\pi/4)
\eeq
 $\theta_{23}=\phi'+\theta$ in the above equation. The mixing angle
$\theta_{12}$
introduced by the symmetry breaking parameter $\tilde{m}$ has two
important effects. It causes a departure from the exact bi-maximal mixing
obtained in the symmetric limit and it leads to splitting among the 
degenerate neutrinos radiatively. Both these features go in the right 
direction in 
solving
the solar neutrino problem.  

In order to evaluate this splitting 
using eq.(\ref{rg}), one needs to express the neutrino mass matrix at $M_X$ in 
the charged lepton mass basis. This can be done using eqs.(\ref{kmx}),
\beq
\label{specific}
M_\nu(M_X)=m\left( \ba{ccc} 
-c_{23}\sin 2\theta_{12}&-c_{23}\cos 2\theta_{12}&s_{23}c_{12}\\
-c_{23}\cos 2\theta_{12}&c_{23}\sin 2\theta_{12}&-s_{23}s_{12}\\
s_{23}c_{12}&-s_{23}s_{12}&0\\
\ea \right) \eeq
where $c_{23}\equiv \cos(\phi'+\theta)\;\;;s_{23}\equiv \sin(\phi'+\theta)$.
In the absence of the symmetry breaking term $\tilde{m}$, $\theta_{12}=0$ 
and the above
matrix has a $L_e-L_\m-L_\tau$ symmetry. Its evolution using eq.(\ref{rg})
does not 
lead to any splitting. With non-zero $\theta_{12}$, one finds using 
eq.(\ref{rg})
\beq
\Delta_S\equiv m_{\nu_2}^2-m_{\nu_1}^2\approx -4 \d_\m m^2c_{23}\sin
2\theta_{12} 
\eeq
Where $\d_\m$ is radiative correction defined in eq.(\ref{deltas}). We have 
neglected 
the electron Yukawa couplings in the above equation. Note that due to the 
specific
texture of eq.(\ref{specific}), the splitting is determined by the $\m$ Yukawa
couplings rather 
than the more dominant tau coupling. With $\Delta_A\equiv m^2$, one obtains,
\beq
\label{diff}
{\Delta_S\over \Delta_A}\sim -{2.7\; \; 10^{-7}\eV^2c_{23}\sin 2
\theta_{12}\over
\cos^2\beta}\eeq

The evolution from $M_X$ to the low scale also affects the leptonic
mixing. This can be determined  after evolving  eq.(\ref{specific}) to the
low scale. One finds that if the leading terms corresponding to only
$\tau$ Yukawa couplings are kept then the
mixing matrix at low scale $\m$ remains formally the same as
eq.(\ref{kmx}): 
\beq
\label{kmu}
K(\m)=R_{12}(\theta_{12})R_{23}(\theta_{23}(\m)) R_{12}(\pi/4) \eeq
where 
$$\tan\theta_{23}(\m)=\tan\theta_{23} (1+\d_\t)\; .$$
It follows that the 
RG evolution does not change the mixing matrix appreciably compared to its form
eq.(\ref{kmx}) at a high scale $M_X$. 

Phenomenological implications of eqs.(\ref{diff},\ref{kmu}) depend upon
the value of $\theta_{12}$ which is determined by the symmetry breaking
parameter $\tilde{m}$. We take it as a free parameter and consider two
interesting extremes corresponding to very small and large $\theta_{12}$
respectively. For small $\theta_{12}$, the leptonic mixing matrix
eq.(\ref{kmu}) has a nearly bi-maximal form. The solar mass scale in
eq.(\ref{diff}) could  lie in the range corresponding to vacuum
oscillation solution. This happens for  moderately large value of
$\tan\b$ which is determined by $\theta_{12}$. More specifically, one
needs
${\sin 2\theta_{12}\over \cos^2\b}\sim 10^{-1}$ in order to obtain 
the solar scale around $\Delta_S\approx 10^{-11}\eV^2$. 

Recent observations \cite{nu00} of neutrino energy spectrum and the day
night
asymmetry are found to be less favourable 
for the vacuum and small angle MSW (SAMSW) solutions respectively.
In contrast, the LAMSW solutions including the one with the low $\Delta_S$
are found to be allowed. One can obtain this  LOW MSW solution  for
somewhat larger value
of $\theta_{12}$ in our case. The mixing matrix (\ref{kmu}) departs from
purely bi
maximal mixing in this case:
\beq
\label{fmix}
K(\mu)=\left( \ba{ccc}
{c_{12}-s_{12}c_{23}^\m\over \sqrt{2}}&
{c_{12}+s_{12}c_{23}^\m\over \sqrt{2}}&s_{12}s_{23}^\m\\
{-s_{12}-c_{12}c_{23}^\m\over \sqrt{2}}&
{-s_{12}+c_{12}c_{23}^\m\over \sqrt{2}}&c_{12}s_{23}^\m\\
{s_{23}^\m\over \sqrt{2}}&-{s_{23}^\m\over \sqrt{2}}&c_{23}^\m \\
\ea \right) \eeq
where $c_{23}^\m=\cos \theta_{23}(\m);s_{23}^\m=\sin\theta_{23}(\m)$.

The above matrix has the correct form to simultaneously solve the solar
and atmospheric neutrino anomalies without violating the constraint from
CHOOZ. The latter constraint does not allow  $\theta_{12}$ to be very
large as it needs 
$$ K_{13}(\m)=s_{12}s_{23}^\m \leq 0.2$$
The data on atmospheric neutrinos require
$$ \sin^2 2\theta_A=4 K_{\m 3}^2(1-K_{\m3}^2)=4 s_{23}^{\m\;2}c_{12}^2
(1-c_{12}^2s_{23}^{\m\;2})\approx 0.84-1$$
The MSW LOW  solution is obtained for
\beqa
\label {range}
\Delta_S&\approx& 
(7-20)\times 10^{-8}\eV^2 \nonumber \\
\sin^2 2\theta_S&\equiv& 4
K_{e1}^2K_{e2}^2=(c_{12}^2-s_{12}^2c_{23}^{\m\;2})^2\approx 0.68-0.98.
\eeqa
Ranges in values of $\theta_{12},\theta_{23}(\m)$ and $\tan\b$ exist which
satisfy
the last three equations simultaneously and lead to the solar scale
in the range  required for the MSW LOW solutions. e.g.
$$ s_{12}\sim 0.28 ; \;\;\;\;\; s_{23}^\m=c_{23}^\m\sim {1\over
\sqrt{2}},\;\;\;\;\;\tan\beta\sim 20
$$
satisfy the CHOOZ constraint and imply 

$$\sin^2 2\theta_S\sim 0.77;\;\;\;\;\;\sin^2 2 \theta_A\sim 0.99;\;\;\;\;
\Delta_S\sim 1.6\times 10^{-7}\eV^2$$
when $\Delta_A\sim 4 \times 10^{-3}\eV^2$.
Since the mass splitting in the model is governed by the $\m$
Yukawa coupling, it is relatively small and one needs to have large
$\tan\b$ in order to obtain the scale relevant for the LOW solution.
\section{Summary}
Solution to the neutrino anomalies, specifically the solar neutrino deficits,
may require scales as small as $\Delta_S\sim 10^{-7}-10^{-11}\eV^2$. It is
interesting to 
suppose that such a scale is generated radiatively. The most economical 
possibility in this context 
would be to assume that the standard model interactions themselves are 
responsible for generating such a small mass difference. This requires specific 
textures for the neutrino mass matrix at high scale. We have discussed in this
paper general conditions
which ensure degeneracy of the masses even after radiative evolution at 
a low scale. It requires specific form, eq.(\ref{genk}) and massless or 
decoupled third neutrino.

We also argued that pseudo Dirac structure originally discussed by Wolfenstein
can arise from a broken discrete symmetry and presented a specific 
example based on discrete symmetry 
in the context of three generations. This example provides a nice realization 
of the 
phenomenologically successful large angle solutions to the neutrino
anomalies.\\ \\
Acknowledgements: We thank A. Dighe and S. Lola for useful comments.\\


\begin{thebibliography}{99}
\bibitem{smir} Different possibilities which simultaneously solve the
solar and atmospheric
neutrino anomalies are reviewed in A. Yu. Smirnov, Talk given at the
International
WEIN Symposium: A Conf. on Physics Beyond Standard Model (WEIN98),    
Santa Fe, June 1998 (hep-ph/9901208); S. M. Barr and I.
Dorsner, hep-ph/0003058;
Anjan S. Joshipura, Pramana, {\bf 54 } (2000) 119.
\bibitem{pdt} G. Dutta and A. S.
Joshipura,\pr{D51}{95}{3838} (hep-ph/9405291); Y. Nir, Journal of High
Energy Physics {\bf 0006 }(2000) 039 ( hep-ph/0002168);
D. Chang and O. C. Kong, hep-ph/9912268;
A. Geiser, hep-ph/9901433
\bibitem{rad1} S. Doi et al, \pr{D30}{84}{626}; Anjan S. Joshipura and
S. D. Rindani, Euro. Phys. Journal, {\bf C14}
(2000) 85 (hep-ph/9811252); W. Grimus and H. Neufeld, hep-ph/9911465; 
S. Lavoura, hep-ph/0005321, T. Kitabayashi and M. Yasue, hep-ph/0006014
\bibitem{rad2} E. Ma, hep-ph/9812344; \prl{83}{99}{2514} (hep-ph/9909249);
R. Adhikari, E. Ma and G. Rajasekaran, hep-ph/0004197;
E. J. Chun and S. K. Kang, hep-ph/9912524.
\bibitem{wolf} L. Wolfenstein, \np{B186}{81}{147}
\bibitem{petcov} S. T. Petcov, \pl{110B}{82}{245}; S. T. Petcov and
C. N. Leung,\pl{125B}{83}{461}
\bibitem{el} K. S. Babu, C. N. Leung and J. Pantaleone, 
\pl{B139}{93}{191}; P. H. Chankowski and Z. Pluciennik,\pl{B316}{93}{312};
M. Tanimoto,\pl{B360}{95}{41}; J. Ellis et al, {\it Eur. Phys. J.} {\bf C9}
(1999) 310; J.
Ellis and S. Lola,\pl{B458}{99}{389}; A. Casas, J. R. Espinosa, A. Ibarra
and I. Navarro,\np{B556}{99}{3}; JHEP {\bf 9909}(99) 015; {\it Nucl.
Phys.} {\bf B569}(2000)
82; K. R. S. Balaji et al, {\it Phys. Rev. Lett.} {\bf 84} (2000)
5034; {\it Phys. Lett.} {\bf  B481} (2000) 33; S. Lola, hep-ph/0005093.
\bibitem{bim} V. Barger, S. Pakvasa, T. J. Weiler and K. Whisnant,
\pl{B437}{98}{107} (hep-ph/9806387); F. Vissani, hep-ph/9708483.
\bibitem{bim1} R. Barbieri et al, \pl{B445}{99}{239}; A. S. Joshipura,
\pr{D60}{99}{053002}; G. Altarelli and F. Fruglio, hep-ph/9905536; A.
Ghosal,
hep-ph/0004171, hep-ph/9905470; Q. Shafi and Z. Tavartkiladze,
hep-ph/0002150;
R. N. Mohapatra et al, Phys. Lett. {\bf 474} (2000) 355; R. Mohapatra and
S. Nussinov,
\pr{D60}{99}{013002} and \pl{B441}{98}{299}; S. Davidson and
S. F. King,\pl{B445}{98}{191};
C. H. Albright and S. M. Barr, \pl{B461}{99}{218}; A. S. Joshipura
and S. D. Rindani, \pl{B464}{99}{239}; M. Masahisa et al, hep-ph/0005147.
\bibitem{brs} R. Barbieri, G. Ross and A. Strumia, hep-ph/9906470
\bibitem{nu00}Talk by Y. Suzuki at XIX Int. Conf. on Neutrino Physics
and Astrophysics, http://ALUMNI.LAURENTIAN.CA/www/physics/nu2000.
\end{thebibliography}
\end{document}